\documentstyle[12pt,epsf]{article}

\begin{document}
\baselineskip=18pt
\pagestyle{empty}
\vspace{-1mm}
\begin{flushright}
{cond-mat/9806355\\
JLC-98-3}
\end{flushright}
\vspace{5mm}
\begin{center}
{\large{\bf Quenched Randomness at First-Order Transitions$^{\dag}$\\}}
\vspace{10mm}
{\bf John Cardy\\}
\vspace{5mm}
{\em Department of Physics\\
Theoretical Physics\\1 Keble Road\\Oxford OX1 3NP, UK\\}
\vspace{3mm}
{\em \& All Souls College, Oxford\\}
\end{center}
\vspace{5mm}
\begin{abstract}
A rigorous theorem due to Aizenman and Wehr asserts that there can be no
latent heat in a two-dimensional system with quenched random impurities.
We examine this result, and its possible extensions to higher
dimensions, in the context of several models. For systems whose pure
versions
undergo a strong first-order transition, we show that there is an
asymptotically exact mapping to the random field Ising model, at the
level of
the interface between the ordered and disordered phases. This
provides a physical explanation for the above result and also implies a
correspondence between the problems in higher dimensions, including
scaling relations between their exponents. The particular example of the
$q$-state Potts model in two dimensions has been considered in detail by
various authors and we review the numerical results obtained for this
case. Turning to weak, fluctuation-driven first-order transitions, we
describe analytic renormalisation group calculations which show how the
continuous nature of the transition is restored by randomness in two
dimensions.
\end{abstract}
\begin{center}
{$^{\dag}$ Invited talk to be presented at STATPHYS 20, Paris, July 1998.}
\end{center}
\newpage

\pagestyle{plain}
\setcounter{page}{1}
\setcounter{equation}{0}
\section{Introduction}
The effect of quenched random impurities which couple to the local energy
density on the critical behaviour of a system whose pure version undergoes 
a continuous transition
is well understood in terms of the Harris criterion \cite{Harris}: if
the specific heat exponent $\alpha$ of the pure system is negative,
such impurities do not change the
qualitative nature of the transition or its universality class, while in
the opposite case there is expected to be different behaviour described
by a new random fixed point of the renormalisation group (RG). Given
the volume of literature devoted to this subject, it is rather
surprising to find relatively few studies of the effects of such
impurities on systems whose pure versions undergo a first-order
transition, despite the ubiquity of such systems in nature. A naive
extension of the Harris criterion, based on the observation that the
effective value of $\alpha$ is unity at a first-order transition, might suggest
that randomness is always strongly relevant. On the other hand,
first-order behaviour is accompanied by a finite correlation length and
is commonly assumed to be stable under small perturbations. Following
early work by Imry and Wortis \cite{IW}, Aizenman and Wehr \cite{AW}
and Hui and Berker \cite{HB} showed that the role of dimensionality $d$
is crucial. In fact the first authors proved a rigorous theorem which
states that for $d\leq 2$ the Gibbs state is always unique, for
arbitrarily small but non-zero concentration of impurities. This means
that there can be no phase coexistence at the transition and hence no
latent heat. 

This result raises a number of important questions: (a) what is its
physical mechanism; (b) is the continuous transition accompanied by a
divergent correlation length; (c) if so, what are the (presumably)
universal critical exponents characterising the transition; and (d) what
happens for $d>2$? Initial Monte Carlo studies of (c) \cite{CFL,Dom}
suggested that the critical behaviour of all such systems might be strongly
universal, exhibiting Ising-like exponents independent of the
underlying symmetry of the model. This received some theoretical
support \cite{Kardar}. However, more recent numerical studies (to be
described below) support earlier theoretical results \cite{Ludwig} in
suggesting that the critical behaviour of, for example, the random bond
$q$-state Potts model depends continuously on $q$, both in the region
where the pure model has a continuous transition, and where it is
first-order. The aim of this talk is to summarise some of these 
recent developments, and to provide at least partial answers to the
above questions.

\section{Mapping to the random field Ising model}
First, I want to describe a mapping \cite{CJ} of this problem to the 
\em random field \em Ising model (RFIM), which is asymptotically valid for very
strong first-order transitions, and which provides a physical
explanation of the Aizenman-Wehr result \cite{AW}, as well as a
prediction for what happens when $d>2$. 

Consider a pure system at a thermal first-order transition point. There
will be coexistence between a (generally unique) disordered phase and
the (generally non-unique) ordered phases. The internal energies 
$U_1$ and $U_2$ of
these two kinds of phase will differ by the latent heat, $L$.
Now consider a interface between the disordered phase and one of the 
ordered phases, with surface tension $\sigma$ (measured in units of $kT_c$).
If $\sigma$ is large, there will be very few isolated bubbles of the
opposite phase above or below the main interface. Its equilibrium
statistics may therefore be
described by a free energy functional equal to its area multiplied by
$\sigma$. Let us compare this with the interface between the two
\em ordered \em phases of an Ising model, with spins taking the values
$\pm1$, at low temperatures. Once
again, there will be few bubbles of the opposite phase, and the bare
interfacial tension will be $\sim 2J$, where $J$ is the reduced exchange
coupling. In the limit when $\sigma\sim 2J$ is large, these
interfacial models are therefore identical\footnote{This is not the case when
bubbles of the opposite phase are included. For example, regions of
ordered phase appearing in the disordered phase are counted with
the degeneracy factor of the ordered phases, as compared with a factor
of unity for the Ising case.}

Now consider the effects of randomness in these two models, in
the first case random bonds, coupling to the local energy density, and
in the second random fields, coupling to the local magnetisation.
In the RFIM, these are accounted for by adding a term $\Sigma_{r>}h(r)-
\Sigma_{r<}h(r)$, where $h(r)$ is the local random field, and in the two
terms $r$ is summed respectively above and below the instantaneous
position of the interface. For random impurities of local concentration
$\delta x(r)$ we have, similarly,
$U_1\Sigma_{r>}\delta x(r)+U_2\Sigma_{r<}\delta x(r)$. Apart from a constant
independent of the position of the interface, this may be written as
$L$ times $\Sigma_{r>}\delta x(r)-\Sigma_{r<}\delta x(r)$, exactly the
same form as for the RFIM. 

We may therefore set up a dictionary between these two cases, in which
the thermal variables of the random bond system are related to the
magnetic variables of the RFIM:
\begin{eqnarray*}
\sigma/kT_c&\longleftrightarrow&J/kT\\
(L/kT_c)\,x&\longleftrightarrow& h_{RF}/kT\\
(T-T_c)\,L&\longleftrightarrow&H\cdot M,
\end{eqnarray*}
where the last relation is between the fields $(T-T_c)$ and a \em
uniform \em magnetic field $H$ which respectively distinguish between
the two phases. Although the above mapping may seem ill-defined in its
use of the local energy density as a kind of order parameter, it may be
made completely explicit, for example, for the $q$-state Potts model
through the mapping to the random cluster model, where $\sigma\sim
L\sim\ln q$ for large $q$ \cite{CJ}.

The interfacial model of the RFIM has been studied extensively
\cite{RFIMinterface} and, in particular, RG equations have been derived
which are asymptotically exact in the limit of low temperatures and
weak randomness, just where our mapping is valid. For $d=2$ the variable
$h_{RF}/J\sim xL/\sigma$ is (marginally) relevant, and, just as this
destroys the spontaneous magnetisation for the RFIM, the latent heat
vanishes for the random bond case, in accord with the Aizenman-Wehr
theorem \cite{AW}. For the RFIM, the RG
trajectories flow towards a paramagnetic fixed point at which the
correlation length $\xi\sim\exp({\rm const}(J/h_{RF})^3)$ is finite, but
note that this is outside the region where the mapping to the random
bond problem is valid. We cannot conclude, therefore, anything about the
nature of the latter's true critical behaviour from this argument. In fact,
numerical and other analytic studies indicate that the actual
correlation length is divergent, and hence $\xi$ is merely a crossover 
length in this case.

The predictions are more interesting for $d>2$, when the RFIM exhibits a
critical fixed point at $kT/J=0$ and a finite value of $h_{RF}/J$.
\begin{figure}
\centerline{
\epsfxsize=3in
\epsfbox{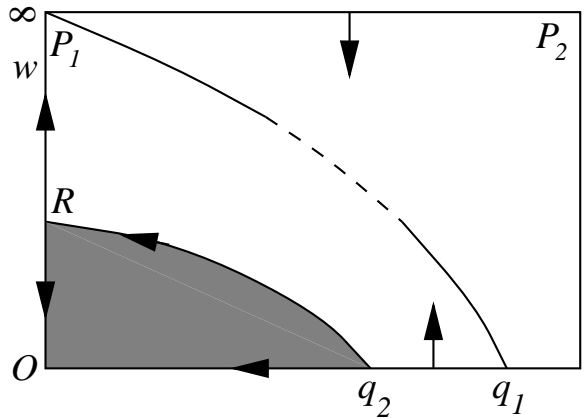}}
\caption{Critical surface for $d>2$, constructed by analogy with the
phase diagram of the RFIM. The axes are $(1/\sigma,x)$ in the
notation of this talk. The shaded region is first-order,
bounded by a line of tricritical points, along which the RG flows go
towards the fixed point at $R$. The upper boundary is the percolation
limit, and the partially dashed curve is a conjectured line of fixed
points describing the continuous random critical behaviour.}
\label{fig1}
\end{figure}
As shown in Fig.~\ref{fig1}, there is now a region in which the
spontaneous magnetisation (latent heat) is non-vanishing as the sign of
the uniform field ($T-T_c$) is changed. This region is bounded, in the
case of the RFIM by the critical curve, close to which the critical
behaviour is determined by the zero-temperature fixed point. Similarly,
the first-order region of the random bond system will be bounded by a
line of \em tricritical \em points, above which (presumably) the transition
becomes continuous. Since the fixed point occurs in the region where the
mapping between the two systems is valid, we may infer some of the
tricritical exponents from those of the RFIM \cite{CJ}. For example, the latent
heat should vanish as $(x_c-x)^\beta$, where $\beta$ is the usual
magnetisation exponent of the RFIM. Similarly, the correlation length on
the critical surface should diverge as $x\to x_c$
with the usual exponent $\nu$ of the RFIM. But the behaviour for 
$T\not=T_c$ is related to the \em magnetic \em properties of the RFIM, and is
complicated by
the fact that the temperature at the RFIM fixed point is dangerously
irrelevant, with an RG eigenvalue $-\theta$ which is responsible for
the violation of hyperscaling $\alpha=2-(d-\theta)\nu$ in that model. 
As a result, for example, the correlation length in the random bond
model at $x=x_c$ diverges as $(T-T_c)^{1/y}$, with 
$y=d-\theta-\beta/\nu$.

\section{Results for the Potts model.}
\begin{figure}
\centerline{
\epsfxsize=4in
\epsfbox{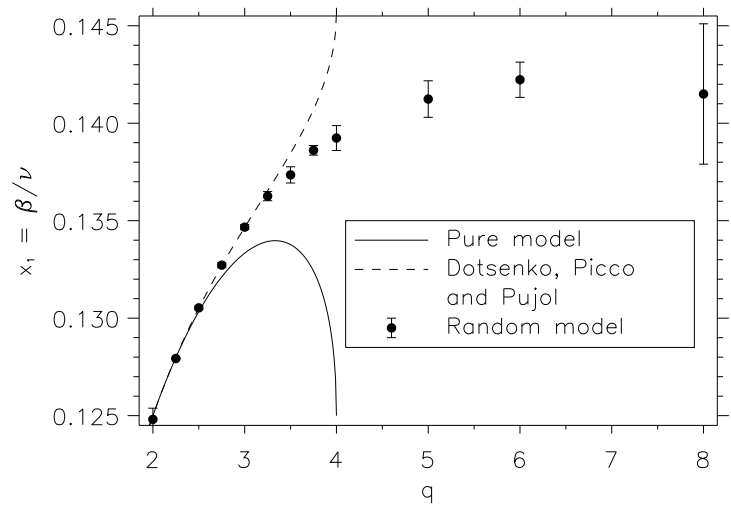}}
\caption{Measured values for the magnetic exponent $\beta/\nu$ of the
random-bond $q$-state Potts model for $d=2$, from Refs.~(9,12). The solid
curve is the exact value for the pure model for $q\leq4$, and the
other is the extrapolated $(q-2)$-expansion of Dotsenko et al. [14].}
\label{fig2}
\end{figure}

As is well-known, the pure $q$-state Potts model undergoes a first-order
transition for $q>4$ in $d=2$, and it is a relatively simple system to
study numerically in the random bond case. By choosing a suitable
distribution of randomness, one can fix the model to be self-dual so that
the critical point is determined exactly. One method, which is very
effective for pure two-dimensional critical system, is to study the
finite-size scaling behaviour of the eigenvalues $\Lambda_i$ of the
transfer matrix in a strip of width $N$. By conformal invariance
\cite{JCconf}, these
are related to the scaling dimensions $x_i$ by 
$2\pi x_i/N\sim\ln(\Lambda_0/\Lambda_i)$ as $N\to\infty$. For the Potts
model, it is possible \cite{JC} to write the transfer matrix in the so-called
connectivity basis \cite{BN}, allowing $q$ to appear as an easily tunable
variable.  In the random case, however, the transfer matrices for
different rows do not commute, and the role of the $\Lambda_i$ is
taken by the Lyapunov exponents which govern the average growth rate
of the norm of vectors under the action of the transfer matrices.
The asymptotic behaviour of the correlation functions is related to
these exponents as for the pure case. 
In order to use conformal invariance, it is necessary to assume
translational invariance which is only recovered after quenched
averaging. However, the Lyapunov exponents themselves are not
self-averaging, only their logarithms. This is related to the phenomenon
of multi-scaling, whereby the average of the $p$th power of the
correlation function is governed by an exponent $x^{(p)}$ which is not
linear in $p$. Since this occurs in most random systems, however, I shall
not treat the problem in detail here. It still proves possible, effectively
by measuring the
whole distribution of the $\ln \Lambda_i$, to extract the decay of the
average correlation function, and hence, for example, the magnetic
exponent $x_1\equiv\beta/\nu$. The results from our study \cite {JC}
are shown in 
Fig.~\ref{fig2}. We see that the value of $x_1$ appears to increase
steadily with $q$. For $q<3$ it agrees with the analytic
$(q-2)$-expansion \cite{Ludwig,Dot}. 
There appears to be no break at $q=4$ where the pure
transition becomes first-order. Our results disagree with earlier Monte
Carlo work \cite{CFL} for $q=8$ where a number close to the Ising value of
$\frac18$ was reported. However, more recent Monte Carlo results by
Picco \cite{Picco} find $x_1=0.150-0.155$ for $q=8$ (and $0.185\pm0.005$
for $q=64$, while Chatelain and Berche \cite{CB} 
report $x_1=0.153\pm0.003$ for $q=8$.
The small discrepancy with our results may be explained by a
careful study of the crossover behaviour which shows that stronger
randomness (beyond the reach of our methods) must be considered
as $q$ increases \cite{Picco}.

All studies report a thermal exponent $\nu\approx1$. This approximately
saturates the lower bound proved by Chayes et al \cite {Chayes}. It
has been suggested \cite{Davis} that, in general, this might be a result of the
averaging procedure and that in some random systems 
the `true' value of $\nu$ might be less than unity. However, it can
shown that this is not the case here.

\section{Analytic results for weak first-order transitions.}
\begin{figure}
\centerline{
\epsfxsize=5in
\epsfbox{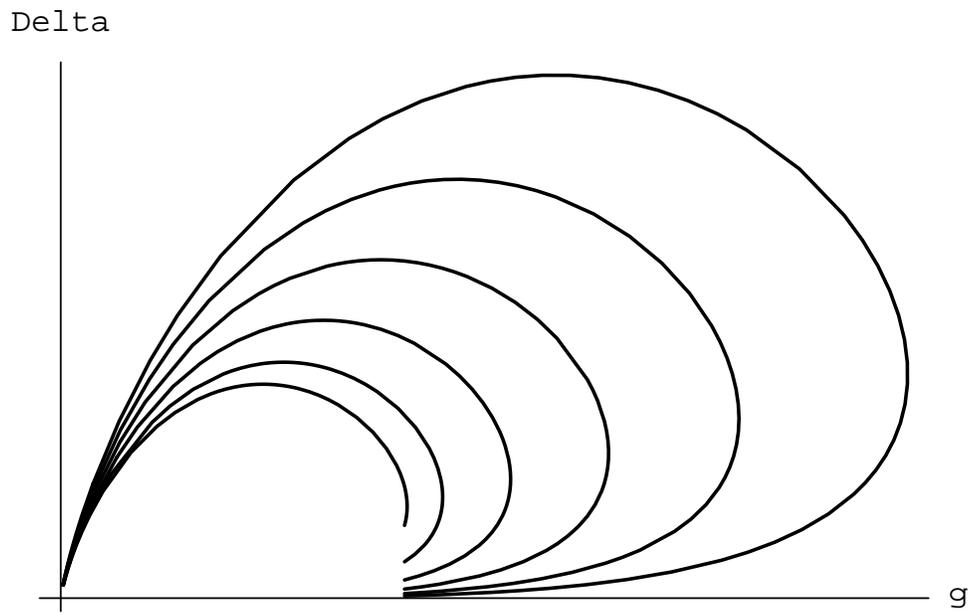}}
\caption{RG flows in the critical surface for $N$ coupled Ising models,
from Ref.~(21).
For non-zero randomness the trajectories curve back towards the
decoupled pure fixed point.}
\label{fig3}
\end{figure}

Although the $q$-state Potts model is relatively simple to analyse
numerically, like most other first-order transitions it is difficult to
study in any kind of perturbative RG approach, since the randomness is
strongly relevant. However, this is
not always the case for systems which exhibit weak `fluctuation-driven'
first-order transitions. An example is afforded by $N$ Ising models
with spins $s_i(r)$,
coupled through their energy densities, with a hamiltonian
\begin{equation}
{\cal H}=-\sum_{r,r'}J(r,r')\sum_is_i(r)s_i(r')+
g\sum_{r,r'}\sum_{i\not=j}s_i(r)s_i(r')s_j(r)s_j(r')
\end{equation}
For uniform couplings $J$, this exhibits a first-order transition when
$N>2$, and in fact, on the critical surface is equivalent to the
Gross-Neveu model, which may be analysed nonperturbatively to show that
the correlation length is $\xi\sim\exp({\rm const}/(N-2)g)$ \cite{GN}. 
Even with
random bonds $J(r,r')$ of strength $\Delta$
the one-loop RG equations may be derived 
By very simple combinatorial methods \cite{JCbook}, to give \cite{RFD}
\begin{eqnarray}
dg/d\ell&=&4(N-2)g^2-8g\Delta+\cdots\label{flows1}\\
d\Delta/d\ell&=& -8\Delta^2+8(N-1)g\Delta+\cdots\label{flows2}
\end{eqnarray}
These flows are illustrated in Fig.~\ref{fig3}.
When $\Delta=0$, $g$ runs away to infinity, in a manner consistent with
the non-perturbative result for $\xi$. When randomness is added,
however, it is marginally relevant, and in fact the flows are quite
similar to those found for the RFIM quoted earlier if we identify 
the interfacial tension $\sim\xi$.
However, in this case, we also see where the flows
end: in this case at the critical
fixed point corresponding to $N$ decoupled pure Ising models. (This is
the only example I know of where the infrared and ultraviolet fixed
points of a set of RG flows are the same.) There have been various
generalisations of this calculation to coupled Potts models \cite{gen}. In most
cases the addition of randomness induces flow towards a critical fixed
point which is perturbatively accessible. In higher dimensions, however,
different outcomes are possible. Adding randomness to models in
$4-\epsilon$ dimensions with cubic anisotropy appears not to change
their fluctuation-driven first-order character \cite{RFD}. However, for impure 
$n$-component superconductors in $4-\epsilon$ dimensions there is a
critical concentration above which the transition becomes continuous
\cite{CBoy}.

\section{Outlook}
There are still many open questions in this relatively little explored
area. A lot more needs to be learned about the nature of the universal
critical behaviour (if indeed it is universal), both in $d=2$ and for
$d>2$ when the randomness is sufficiently strong. In the former case, it
may be possible to solve the problem by conformal field theory methods.
Even the limit of large $q$ appears non-trivial, however. Similar ideas
apply to quantum phase transitions and may have relevance for the
quantum Hall effect. Most importantly, it should be possible to find
experimental systems which realise some of the predictions discussed
above. There are, after all, many three-dimensional examples of
first-order transitions. However, it should be stressed that the
randomness should couple only to the local energy density, not to the
order parameter, otherwise this becomes the random field problem. 
It is also important to ensure that the tricritical
behaviour is driven by the effects discussed and not by some simple
mean-field mechanism (which would give rise to mean-field tricritical
exponents in $d=3$, part from logs.) Finally it is important to
understand the dynamics of these systems: it may be that, like the
random field problem, they are plagued by logarithmically slow time
scales close to the critical point \cite{RFdyn}.

I am grateful to Jesper Jacobsen for his continuing collaboration on
this problem. This work was supported in part by EPSRC Grant GR/J78327.

\end{document}